# Nanoscale Patterning of Surface Nanobubbles

Anayet Ullah Siddique, Rui Xie, Danielle Horlacher, and Roseanne Warren*

Department of Mechanical Engineering

University of Utah

1495 E 100 S, 1550 MEK

Salt Lake City, UT 84112, USA

*roseanne.warren@utah.edu

**Highlights**:

- Surface nanobubble patterning using focused ion beam (FIB) technique.
- Nanobubbles selectively positioned on hydrophobic domains.
- Location of surface nanobubbles regulated by nanostructured surfaces.
- Effect of FIB patterning on morphology of nanobubble including contact angle analyzed.

**Abstract (180 words)**

Surface nanobubbles forming on hydrophobic surfaces in water present an exciting opportunity as potential agents of top-down, bottom-up nanopatterning. The formation and characteristics of surface nanobubbles are strongly influenced by the physical and chemical properties of the substrate. In this study, focused ion beam (FIB) milling is used for the first time to spatially control the nucleation of surface nanobubbles with 75 nm precision. The spontaneous formation of nanobubbles on alternating lines of a self-assembled monolayer (octadecyltrichlorosilane) patterned by FIB is detected by atomic force microscopy. The effect of chemical *vs*. topographical surface heterogeneity on the formation of nanobubbles is investigated by comparing samples with OTS coating applied pre- *vs*. post-FIB patterning. The results confirm that nanoscale FIB-based patterning can effectively control surface nanobubble position by means of chemical heterogeneity. The effect of FIB milling on nanobubble morphology and properties, including contact angle and gas oversaturation, is also reported. Molecular dynamics simulations provide further insight into the effects of FIB amorphization on surface nanobubble formation. Combined, experimental and simulation investigations offer insights to guide future nanobubble-based patterning using FIB milling.

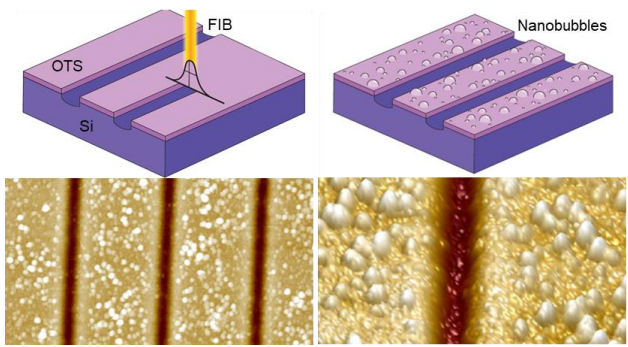

## 1. Introduction

Surface nanobubbles are nanoscale spherical-cap-shaped gaseous domains that form on various surfaces immersed in water [1,2]. To-date, the properties and applications of surface nanobubbles nucleated on homogeneous substrates have been well-studied. Surface nanobubbles tend to form on hydrophobic surfaces, with some reports of nanobubbles on hydrophilic surfaces [3,4]. Applications of surface nanobubbles explored to-date include: surface cleaning enhancement [5,6], mineral flotation [7], wafer-scale graphene transfer [8], and manipulation of no-slip boundary conditions in microfluidic channels [9]. While such applications demonstrate the utility of large-scale, uniform distributions of surface nanobubbles, equally intriguing are the potential applications of spatially-confined nanobubbles. In particular, the ability to control surface nanobubble placement at sub-100 nm length scales opens the door to new methods of top-down/bottom-up nanoscale patterning using surface nanobubbles.

Several prior works have achieved spatial control of nanobubble formation using nanopatterned hydrophobic/hydrophilic surfaces. Agrawal *et al.* demonstrated preferential nucleation of surface nanobubbles on the hydrophobic regions of poly (methyl methacrylate) (PMMA)-polystyrene (PS) block copolymer nanodomains [10]. Hydrophobic-hydrophilic patterns for nanobubble positioning have also been achieved *via*: electron beam-induced deposition (Teflon-carbon nanodomains) [11]; microcontact printing (octadecanethiol-*n*-octadecylphosphonic acid hydrophobic domains) [12]; and electron beam lithography (patterned PMMA films) [13]. The ability for nanobubbles to serve as templates for bottom-up patterning has been demonstrated through, *e.g.*: surface nanobubble-induced PS nanoindents [14,15]; NaCl solution evaporation around surface nanobubbles [16]; convective self-assembly of Au nanoparticles around nanobubbles and droplets [17]; and microporous Cu electrodeposition around bubble templates [18].

This work presents the first demonstration of sub-100 nm, top-down, spatial control of surface nanobubbles using focused ion beam (FIB) milling. FIB milling offers numerous advantages compared to previously demonstrated surface nanobubble patterning approaches, including: 1) higher resolution, reduced proximity effects, and higher throughput *vs.* electron beam lithography [19,20]; 2) direct-write patterning that avoids stamp fabrication, as in microcontact printing; and 3) precise placement of features, unlike self-assembly-based methods such as block copolymer lithography. There are, however, several noteworthy drawbacks of FIB milling that directly impact its compatibility with surface nanobubble patterning. In the process of ion beam exposure, FIB milling creates myriad changes to the underlying substrate, including: amorphization of Si (including up to ~20 nm depth) [21], $Ga^+$ ion implantation [22], and changes in surface roughness [23]. Molecular dynamics (MD) simulations have demonstrated changes in water contact angle [24], in addition to nano-order deformation and hillock structure formation [22]. Further complicating matters is the spatial gradient of such effects arising from the Gaussian beam shape, which spreads beyond the user-defined patterning region [25]. As a result of these complex ion-surface interactions, experimental demonstration is needed to validate the concept of nanoscale FIB-based patterning of surface nanobubbles.

Figure 1a provides an overview of the approach employed here to achieve surface nanobubble patterning using FIB milling. A Si surface is coated with the hydrophobic self-assembled monolayer (SAM) octadecyltrichlorosilane (OTS) (Figure 1a, left inset). Upon exposure to the FIB, the SAM is selectively removed from the substrate. When immersed in water, surface nanobubbles will selectively form in the SAM regions. FIB patterning can induce full or partial monolayer removal or degradation, depending on factors such as beam dose and current density [26,27]. Changes to the underlying Si surface such as ion implantation and amorphization are also expected (Figure 1a, right inset). To distinguish FIB *vs.* non-FIB regions by atomic force microscopy (AFM), it is necessary to provide some depth contrast to the line patterns. To separate ion beam exposure effects from topographical effects, two samples were made for comparison. The first sample follows the intended nanopatterning approach of OTS coating, followed by FIB milling (Figure 1b, top), and is referred to as "hydrophobic/hydrophilic" (HB/HL). The control sample was FIB milled first, followed by OTS coating (Figure 1b, bottom), and is called "hydrophobic/hydrophobic" (HB/HB).

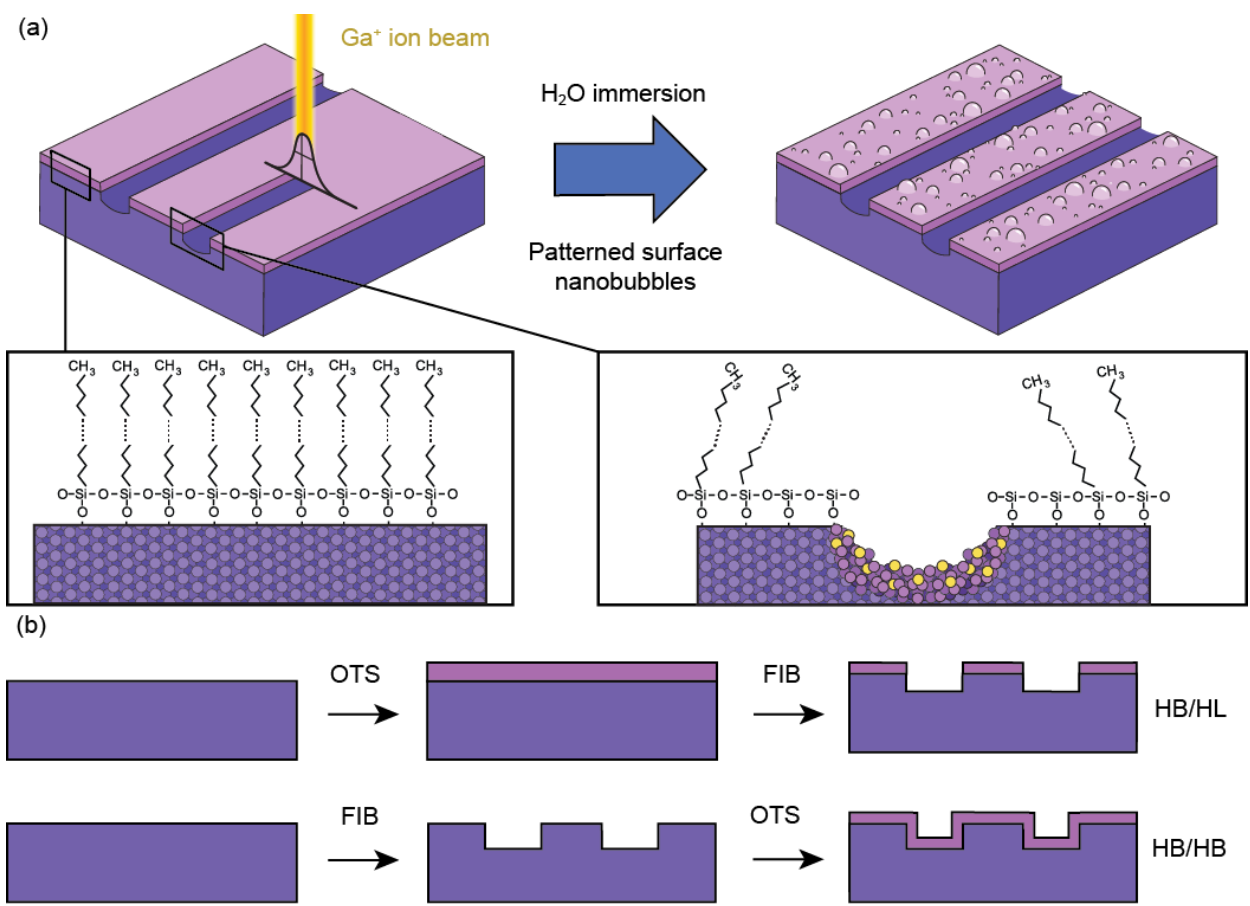

**Figure 1.** (a) Conceptual illustration of FIB-based patterning of surface nanobubbles. A hydrophobic, SAM (OTS) is applied to a Si surface, patterned by FIB milling, then immersed in water. Left inset: structure of the pristine OTS layer. Right inset: FIB milling effects include: full or partial removal of the SAM, $Ga^+$ ion implantation, and Si amorphization. (b) Process flow diagrams for the two types of nanostructured patterns: HB/HL (top) and HB/HB (bottom).

**2. Materials and methods:**

2.1 Materials

OTS (95%) and toluene (anhydrous, 99.8%) were purchased from Fisher Scientific. Acetone (> 99.9%) and ethanol (> 99.8%) were purchased from Sigma Aldrich. Chloroform (≥ 99% ACS Reagent Grade) was purchased from Lab Alley. Ultrapure water (conductivity 18.2 MΩ cm) was obtained from a Milli-Q system (Millipore Corporation, Boston, MA). To avoid contamination, all fluid handling was conducted using glass beakers and syringes cleaned with ethanol and ultrapure water before use. Before each experiment, the AFM liquid cell was rinsed with isopropyl alcohol, ethanol, and ultrapure water, and dried with nitrogen gas.

2.2 Substrate modification

OTS-modified silicon surfaces were prepared by first treating the silicon wafer with oxygen plasma for 2 minutes at 195 W. A 5 mM OTS solution was prepared by dissolving OTS in toluene in a Class 1000 clean room environment. The silicon wafer was immersed in the OTS solution and kept in a sealed container for 24 h. Upon removal from the solution, the OTS-modified wafer was quickly rinsed with chloroform. The wafer was sonicated for 15 minutes each in chloroform, toluene, and acetone to remove unbound OTS molecules. The wafer was then dried under a stream of nitrogen gas and kept in the cleanroom for at least 24 h. Before each AFM experiment, all substrates were sonicated in toluene, acetone, and ethanol in turn for 5 minutes each and dried under a stream of nitrogen gas.

     An FEI Helios Nanolab 650 focused ion beam-scanning electron microscope (FIB-SEM) was used to fabricate the nanostructures. NanoBuilder$^{TM}$ software was used to design the fabricated structures. Single-pixel lines 20 µm long by 75 nm width were milled using a voltage of 30 keV and an ion beam current of 24 pA. HB/HL surfaces were prepared by OTS coating

followed by FIB milling of nanostructures (Figure 1b, top). HB/HB surfaces were prepared by FIB milling of nanostructures, followed by OTS coating (Figure 1b, bottom).

2.3 AFM experiments

AFM measurements were conducted using a Dimension Icon AFM (Bruker, USA). AFM experiments in air were performed using a silicon nitride cantilever (Bruker) with a nominal spring constant of 0.4 N/m and a tip radius of 2 nm. AFM experiments in fluid were performed using a silicon nitride cantilever (DNP-C, Bruker) with a nominal spring constant of 0.24 N/m and a tip radius of 20 nm. ScanAsyst® mode was used in both air and fluid experiments. Before each fluid experiment, the AFM liquid cell was rinsed with isopropyl alcohol, ethanol, and ultrapure water, and dried with nitrogen gas. The fluid cell was first mounted with the AFM head (scanner) and sealed by a translucent silicone O-ring. A drop of Milli-Q water was then added directly to the sample substrate using a glass syringe, and another drop was placed on the liquid cell. When the optical head was lowered, the two droplets coalesced and were squeezed to form a meniscus between the substrate and the liquid cell. Prior to beginning the experiments, the system was left for 30 minutes to reach thermal equilibrium. All AFM experiments were performed at room temperature. Typical scan areas were 2 µm x 2 µm, and the scan time for each image was approximately 8-9 minutes. Bruker NanoScope Analysis 2.0 software was used for AFM image processing.

2.4 Degassed experiments

Degassed fluid AFM experiments were performed using DI water placed inside a desiccator at room temperature for 24 hours under reduced pressure. The time between the degassed water being removed from the desiccator and injected into the fluid cell was as short as possible. The 30 minute equilibration time before imaging was bypassed in degassed experiments.

2.5 Contact angle measurement

The water contact angle was measured using a contact angle goniometer (Model no. 100-00-115, ramé-hart instrument co.). The contact angles of water on the bare silicon wafer and hydrophobized flat OTS coated silicon wafer were measured by the sessile drop method. Each measurement was repeated at least 5 times at different surface locations for each substrate, and the average result was reported.

2.5 MD simulations

MD simulations were implemented in Large-scale Atomic/Molecular Massively Parallel Simulator (LAMMPS). The size of the system was approximately 15 nm x 4.5 nm x 25 nm in the $x$, $y$, and $z$ directions, respectively. The potential for the surface preparations was the three-body Tersoff potential. The substrate was first annealed at 2000 K with NVT ensemble for 2 ns, then relaxed at 2000 K with NVE ensemble for another 2 ns, and then quenched to 300 K at a rate of $10^{12}$ K·s$^{-1}$. Periodic boundary conditions were applied in the $x$, $y$, and $z$ directions for this step. After the amorphous substrate was prepared, the extended simple point charge (SPC/E) water model with fixed bond length and angles was prepared. Water molecules were separated 3.1 Å from each other and mixed with 1.4% or 3% nitrogen molecules at least 1 Å above the substrate as the initial configuration. Periodic boundary conditions were applied in the $x$ and $y$-directions, and the mirror boundary condition was applied in the $z$-direction.

Nanobubble simulations were implemented using the canonical (NVT) ensemble, with temperature controlled by the Nose–Hoover thermostat at 300 K and a timestep of 1 fs. Long-range Coulombic interactions were calculated by the particle-particle particle-mesh (PPPM) method. The Lennard-Jones (LJ) potential with a cutoff value equal to 12 Å was applied for the Van der Waals interaction. Lorentz-Berthelot mixing rules were applied if the species were different, and LJ potential parameters were obtained from [24,28,29].

3. Results and Discussion

3.1 Characterization of nanopatterned surfaces

Figure 2 provides schematic (Figure 2a-c), SEM (Figure 2d-f), and AFM (Figure 2g-l) images of nanostructured surfaces fabricated *via* FIB milling. In Figure 2d, the Si surface is uncoated (no OTS layer) and displays a regular pattern of trenches of width 75 nm and period 500 nm. Figure 2e provides an SEM image of the HB/HL surface, in which FIB milling was completed after OTS coating. Figure 2f provides an SEM image of the HB/HB surface, prepared by OTS coating after FIB milling. The line width and period of the FIB-milled trenches is consistent across all three surfaces (uncoated, HB/HL, and HB/HB), indicating no significant shrinkage of line width occurs with the OTS layer. In all cases, the FIB-milled nanopatterned surfaces are regular and uniform across the surface.

Figure 2g-l provides AFM images of uncoated Si, HB/HL, and HB/HB nanopatterns in the air. While the uncoated Si surface is clear (Figure 2g), small particles are visible on the OTS-modified surfaces (Figure 2h, i). These particles may be caused by small amounts of OTS aggregation during coating. AFM cross sections (Figure 2j-l) confirm the depth of the trenches is ~7 nm for uncoated, HB/HL, and HB/HB surfaces. The trench width of 75 nm is unaffected by the OTS coating, and remains consistent for uncoated, HB/HL and HB/HB surfaces. A step height of ~ 7 nm was chosen to differentiate between hydrophobic and hydrophilic regions during fluid cell AFM imaging. A trench width of 75 nm was chosen so that the fluid cell AFM tip (~ 20 nm radius) can completely enter the trenches.

Unpatterned OTS on Si has a measured water drop contact angle of $110° \pm 2°$ *vs.* $36° \pm 1°$ for pure Si. The measurement confirms that the HB/HL surface provides both chemical and topographical surface heterogeneity, while the HB/HB surface provides only topographical surface heterogeneity.

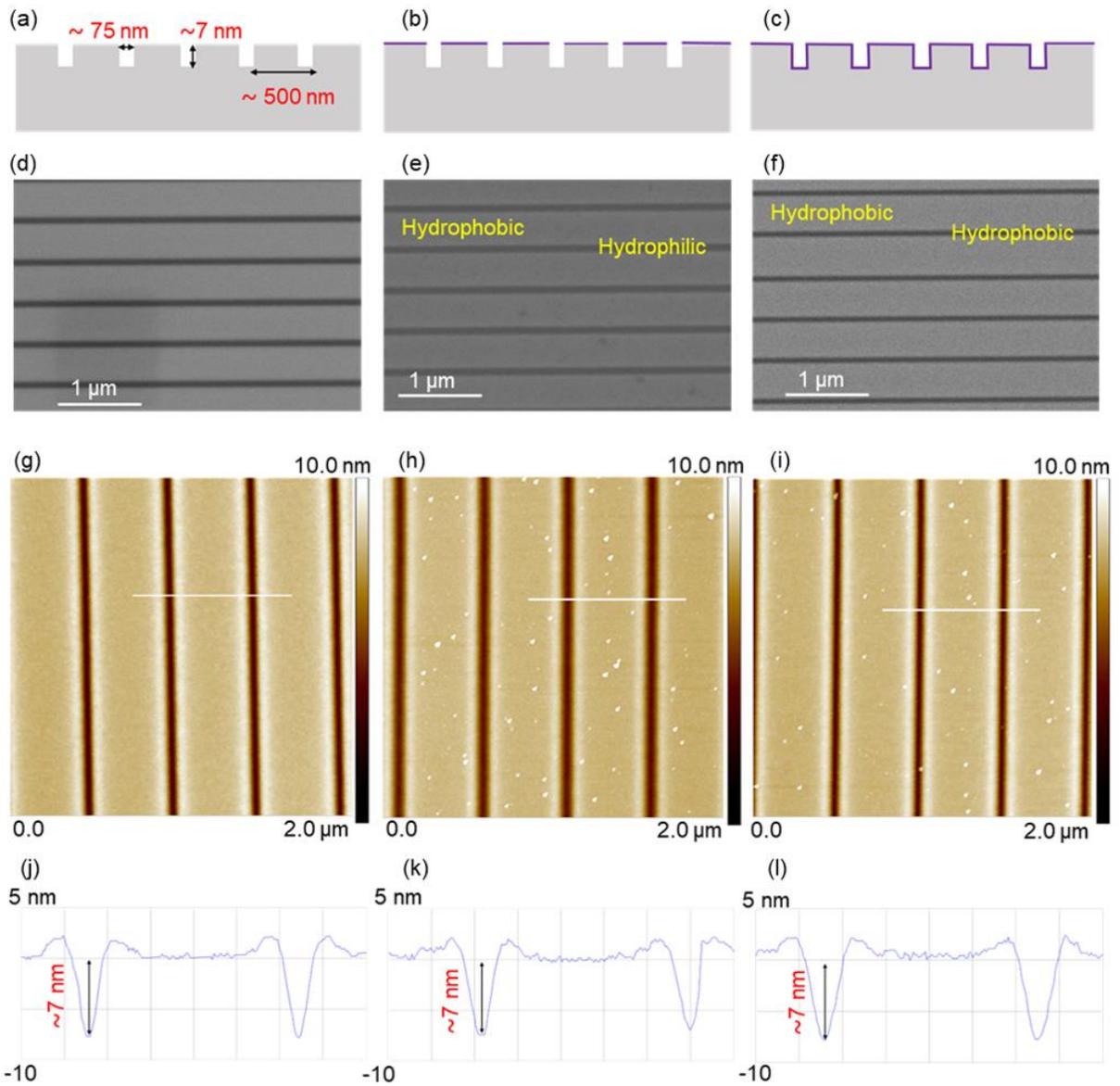

**Figure 2.** (a-f) Schematic and SEM images of: a, d) uncoated Si after FIB milling; b, e) HB/HL nanopatterned surface; and c, f) HB/HB nanopatterned surface. Schematic diagrams indicate surface regions coated with OTS (purple shading). (g-i) AFM images of: g) FIB milled uncoated Si; h) HB/HL; and i) HB/HB surfaces in air. (j-l) Cross-sectional AFM profiles of (g-i), respectively. Profiles are taken along the white lines drawn in (g-i).

3.2 Fluid AFM imaging

Fluid AFM imaging was first conducted on a planar OTS-modified Si substrate to measure the properties of surface nanobubbles formed when a water droplet contacts the hydrophobic surface (Figure 3). The AFM height image (Figure 3a) shows the presence of dense, randomly located features on the OTS-coated surface. Line scans of these features indicate bubble-like profiles of typical height of approximately 10 nm and width of approximately 50 nm (Figure 3b, c). To confirm these features are bubbles and not artifacts of surface preparation or imaging, we performed degassed fluid AFM experiments. The results confirm that the features are indeed bubbles, appearing on OTS-Si only in the presence of dissolved gas, with no surface features visible in the degassed experiment (Supporting Information Figure S1).

Fluid AFM imaging was then conducted on HB/HL and HB/HB nanopatterned substrates to investigate the presence of nanobubbles on the surfaces *vs.* in the trenches of both samples (Figure 4). In the case of HB/HL patterns, surface nanobubbles are visible on the upper, hydrophobic stripes but not in the hydrophilic trenches (Figure 4a-c). For the HB/HB patterns, nanobubbles formed on both the upper stripes and within the grooves between the stripes (Figure 4d-f). Line scans of HB/HL (Figure 4b) and HB/HB (Figure 4e) samples indicate comparable height profiles for bubbles located on the hydrophobic regions of HB/HL and HB/HB surfaces. The features are also analogous to bubble shapes indicated in line scans of unpatterned OTS (Figure 3b, c). Supporting Information Figure S2 provides additional fluid AFM images confirming the presence of nanobubbles in HB/HB trenches and the absence of bubbles in HB/HL trenches.

Table 1 quantifies the presence *vs.* absence of nanobubbles in HB/HB *vs.* HB/HL trenches by comparing "AFM feature density" counts for both samples. While the fluid AFM tip is able to scan within the full width of the trenches, it is not possible to obtain accurate values of surface nanobubble density within the trenches due to uncertainties distinguishing individual bubble features within trenches. A given AFM feature, for example, may consist of one or several bubbles that cannot be resolved with confidence. Feature density counts provided in Table 1 are based on Figure 4 images. The results indicate a clear absence of nanobubbles in HB/HL trenches compared to HB/HB trenches and HB/HL or HB/HB upper stripes. Assuming similar errors in using "AFM feature density" as a proxy for surface nanobubble density in both samples, there is an approximate 83% reduction in bubble feature density for HB/HL surfaces in FIB-milled areas. The results demonstrate the ability to control placement of surface nanobubbles using chemical heterogeneity achieved by FIB milling of OTS.

**Table 1:** AFM feature density for HB/HL and HB/HB patterns.

|  | HB/HL (number/$\mu m^2$) | HB/HB (number/$\mu m^2$) |
| --- | --- | --- |
| Trenches | 3.3 | 20.0 |
| Stripes | 18.2 | 21.2 |

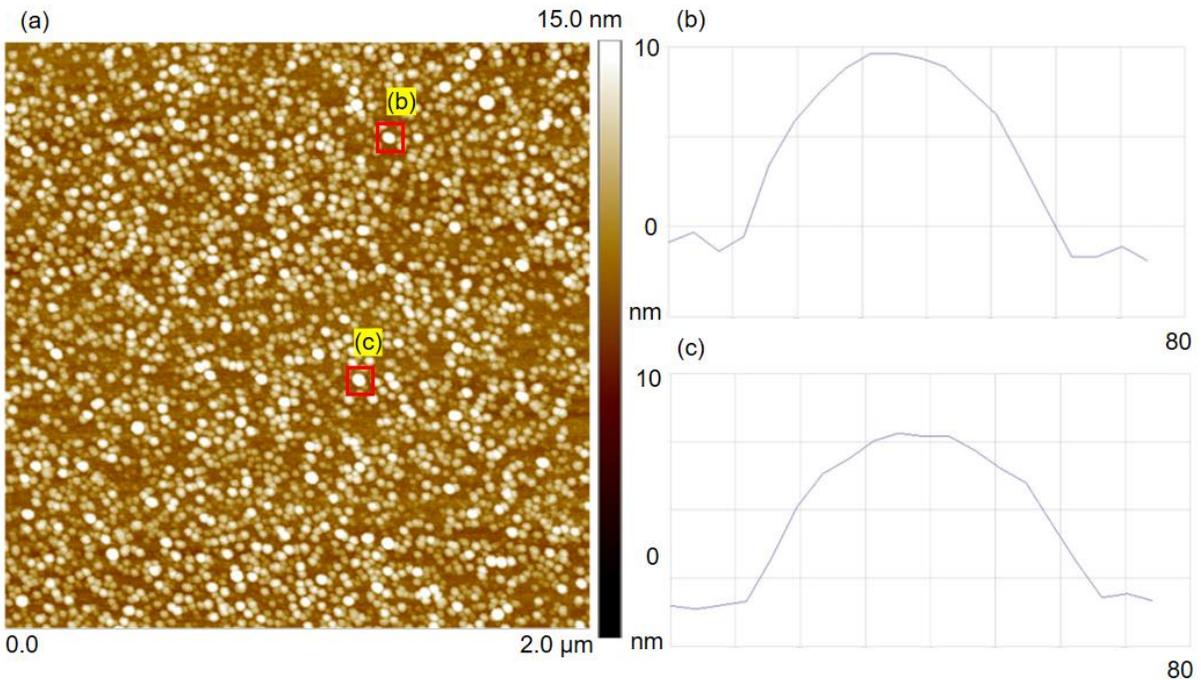

**Figure 3.** (a) Typical AFM height image of nanobubbles formed on the planar OTS coated Si. (b, c) Height profiles of the nanobubbles in (a). Profile location is indicated by a red square in part (a).

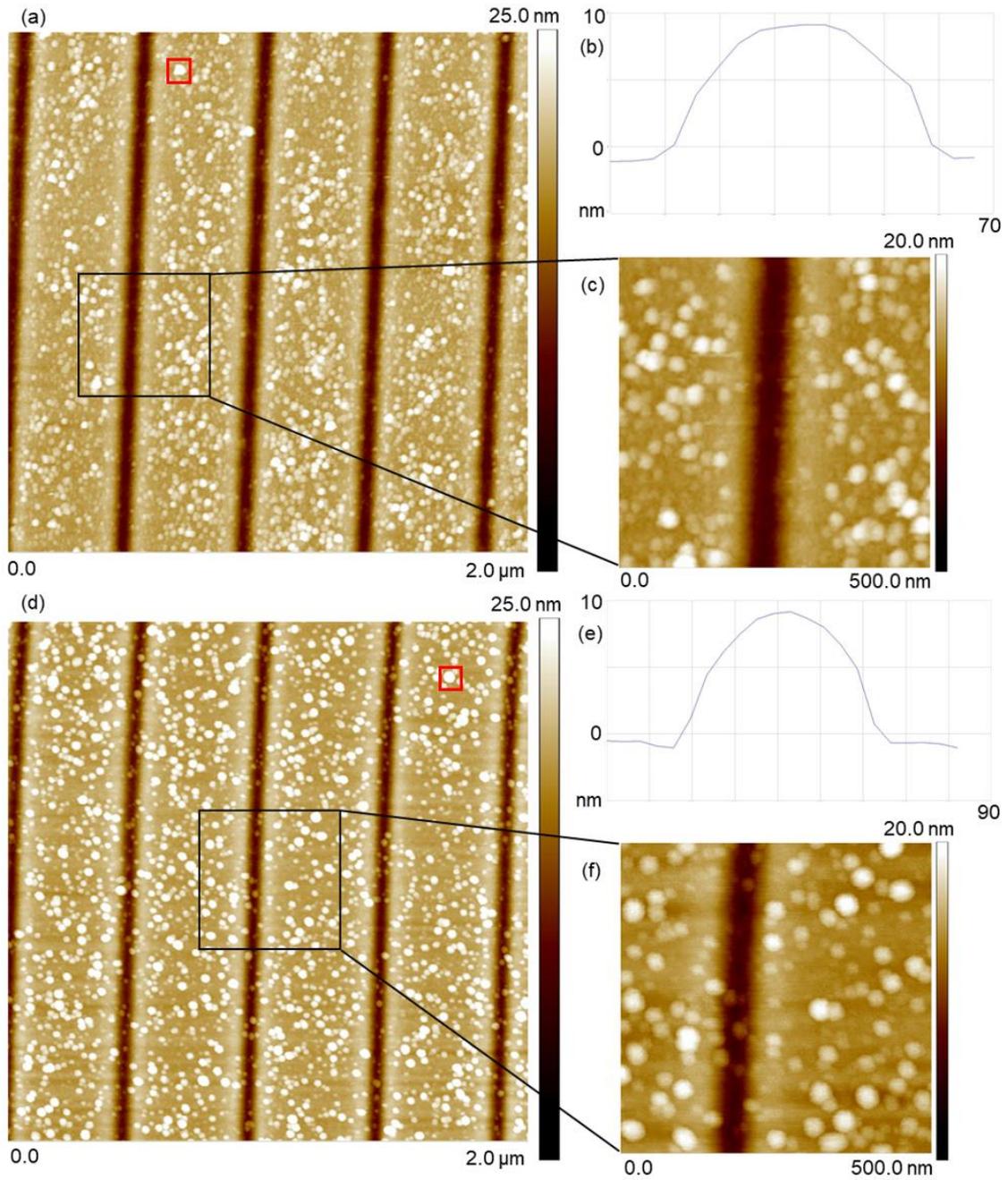

**Figure 4.** Fluid AFM measurements of nanobubbles formed on HB/HL (a-c) and HB/HB (d-f) surfaces. (a, d) AFM height images of nanobubbles on HB/HL (a) and HB/HB (d) surfaces, with a scan size of 2 µm x 2 µm. (b, e) Height profiles of HB/HL (b) and HB/HB (e) nanobubbles. Profile locations are marked with a red square in the corresponding AFM images. (c, f) Zoomed-in image of a single HB/HL (c) or HB/HB (f) trench with a scan size of 500 nm x 500 nm.

3.3 Nanobubble properties

The morphology of nanobubbles present on HB/HL *vs*. HB/HB *vs*. unpatterned OTS surfaces was quantified by comparing bubble height (*H)* and base width (*L*). Figure 5 presents height and base width histograms fit with a Gaussian distribution for unpatterned OTS (Figure 5a, b), HB/HL (Figure 5c, d), and HB/HB (Figure 5e, f) surfaces. Mean height, mean width, size range, and standard deviation values are summarized in Table 2. Due to uncertainties obtaining exact lateral dimensions of nanobubbles bubbles formed within trenches, height and width values are reported for bubble features present along upper surface stripes only. The histograms indicate similar morphologies for nanobubbles on plain OTS *vs*. HB/HL and HB/HB patterned substrates. Statistical analysis was performed to test for significant differences in height and width between bubbles forming on flat OTS surfaces *vs.* nanopatterned surfaces (HB/HL and HB/HB). A paired-sample t-test was used with a significance level (p-value) of 0.05. The average widths of the nanobubbles on the HB/HL and HB/HB substrates were 44.32 and 43.32 nm, respectively, which were not significantly different ($p > 0.05$) from the average width of the nanobubbles on unpatterned OTS (mean = 45.10 nm). No significant difference ($p > 0.05$) was observed between the heights of nanobubbles on HB/HL (mean = 8.12 nm) *vs*. unpatterned OTS (mean = 7.75 nm), or nanobubbles on HB/HB (mean = 7.87 nm) *vs*. unpatterned OTS.

Outcomes of this morphology comparison indicate that there is no measurable change in OTS coating that occurs as a result of the FIB milling process. With the ion beam current (24 pA) and voltage (30 keV) employed here, the OTS layer is effectively removed within the 75 nm line width without altering adjacent monolayer properties. Potential problems with OTS coating degradation or partial removal documented in previous studies and presented in Figure 1 are found not to inhibit the use of FIB milling as a top-down/bottom-up patterning approach with OTS layers on Si.

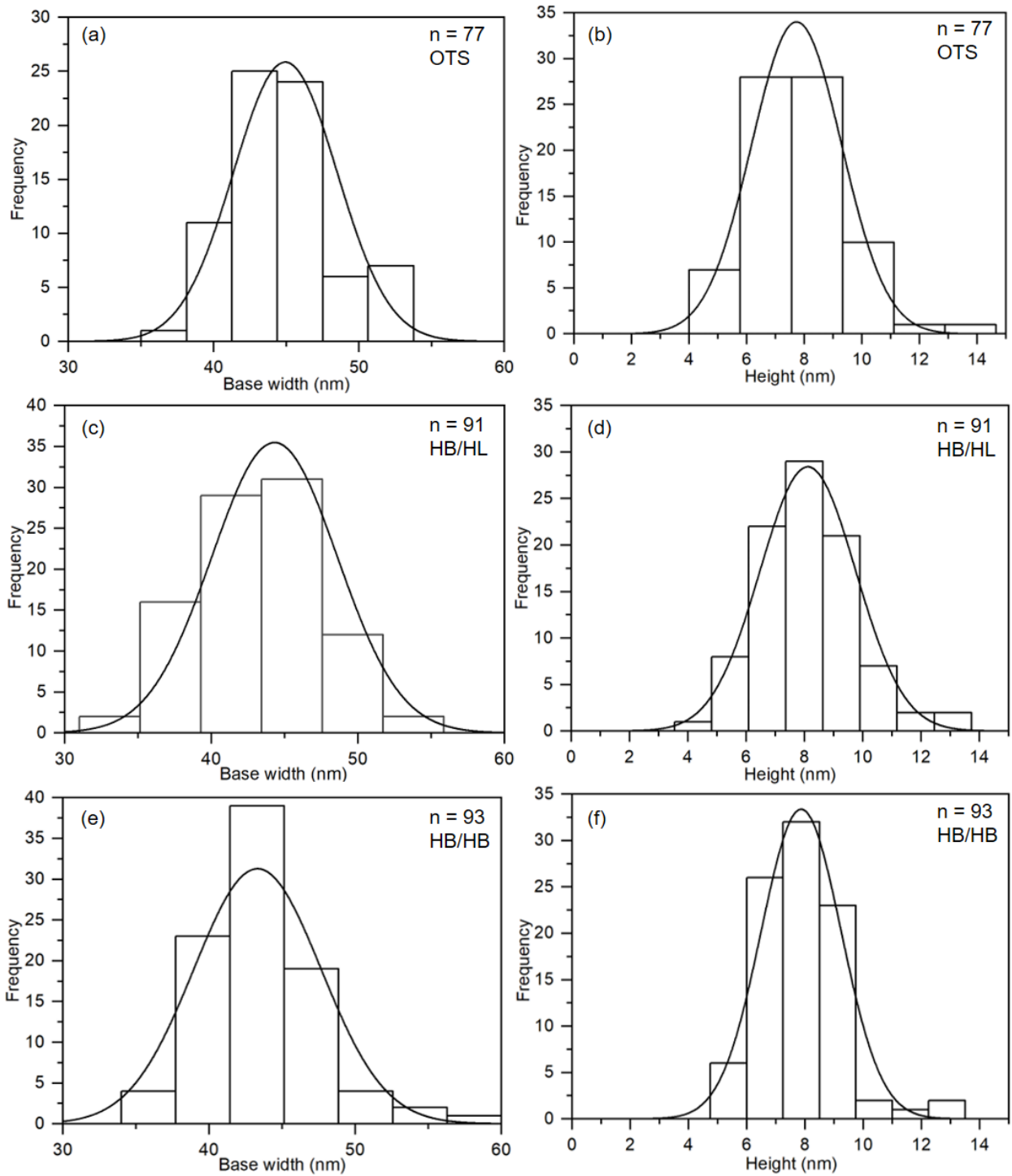

**Figure 5.** Base width (a, c, e) and height (b, d, f) histograms of nanobubbles on flat OTS (a, b), HB/HL (c, d), and HB/HB (e, f) surfaces. Histograms are fit with Gaussian curves.

**Table 2:** Statistical table of the size of nanobubbles forming on different surfaces

| Geometric parameters | | Flat OTS | HB/HL | HB/HB |
|---|---|---|---|---|
| Width, $L$ (nm) | Mean | 45.1 | 44.315 | 43.32 |
| | Range | 38-59 | 34-55 | 35-59 |
| | Stdv | 3.873 | 4.263 | 4.33 |
| Height, $H$ (nm) | Mean | 7.731 | 8.116 | 7.8696 |
| | Range | 4.57-13.1 | 4.08-12.83 | 5.07-13.1 |
| | Stdv | 1.553 | 1.635 | 1.367 |

Additional nanobubble properties of interest include the contact angle *vs.* height, and equilibrium contact angle ($\theta_e$). Figure 6a plots the relationship between contact angle ($\theta$) and height for nanobubbles measured on HB/HL and HB/HB surfaces, with $\theta$ approximated by Equation 1:

$$\theta = tan^{-1}\left(\frac{2H}{L}\right) \quad (1)$$

Nanobubbles produced on HB/HL and HB/HB surfaces have contact angles in the range of 13.4°-28.5°, and 16°–25.5°, respectively, with a combined average contact angle of 20°. The increase in contact angle with height shown in Figure 6a is consistent with previous nanobubble reports [30–32]. Differences in height, width, and contact angle for flat OTS, HB/HL and HB/HB nanobubbles *vs.* literature reports for OTS surfaces are likely due to differences in bubble formation methods, *i.e.* lower dissolved gas content in the water droplet method employed here *vs.* fluid exchange or temperature difference methods [33].

The relationship between of gas oversaturation, $\zeta$, and the equilibrium contact angle can be evaluated using the Lohse-Zhang theory [33,34] as follows:

$$\zeta = \frac{4\sigma}{P_{atm}L} sin\theta_e \quad (2)$$

where $\sigma$ is air−liquid surface tension (0.072 Nm$^{-1}$) and $P_{atm}$ is atmospheric pressure. Figure 6b plots $sin\theta$ vs. $L$ for HB/HB and HB/HL nanobubbles. According to Equation 2, gas oversaturation can be calculated from a linear fit of $sin\theta/L$. The calculated $\zeta$ value corresponds to an air saturation of $\zeta = 8.2$, which is plausible and in line with the value predicted by the Lohse−Zhang theory [34]. According to the theory, nanobubbles must be sustained by oversaturation of at least $\zeta \approx 4-7$ [34,35], which indicates that gas oversaturation is a key factor for the formation of surface nanobubbles on HB/HL and HB/HB surfaces.

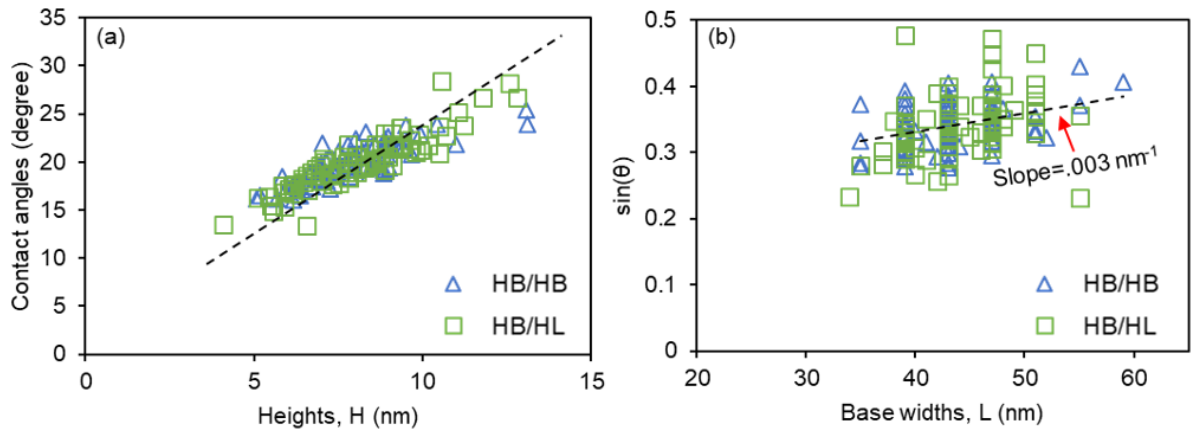

**Figure 6.** (a) Linear dependence of contact angle *vs* height of nanobubbles on HB/HL and HB/HB surfaces. (b) $sin\theta$ vs. nanobubble width $L$ for HB/HL and HB/HB surfaces. The slope of 0.003 nm$^{-1}$ corresponds to a gas oversaturation of $\zeta = 8.2$.

3.4 Effect of Amorphization on Nanobubble Formation on Si

During FIB patterning, crystalline Si can be converted to amorphous Si. To confirm that there is no tendency towards bubble formation on amorphous Si, and to better understand nanobubble formation on the molecular level in the trench, MD simulations were performed on crystalline and amorphous Si surfaces. Figure 7 provides simulation results for (100) Si and amorphous Si substrates after 12 ns. Amorphous Si substrates are designated "100 Amorphous" and "111 Amorphous" in Figure 7 based on the starting single-crystal orientation used to form the amorphous phase in the simulations. For 1.4% $N_2$ concentration, no bubbles formed on either crystalline (Figure 7a) or amorphous (Figure 7b, c) Si after 12 ns. In experiments, the number percentage of nitrogen molecules is much less than 1.4% $N_2$. We therefore conclude that the FIB milling process will not produce a Si surface conducive to nanobubble formation at typical dissolved gas quantities. This confirms the feasibility of FIB-based patterning to selectively define nanobubble and nanobubble-free regions on a Si substrate. Supporting Information Figure S3 and Table S1 provide information on additional scenarios evaluated through simulations.

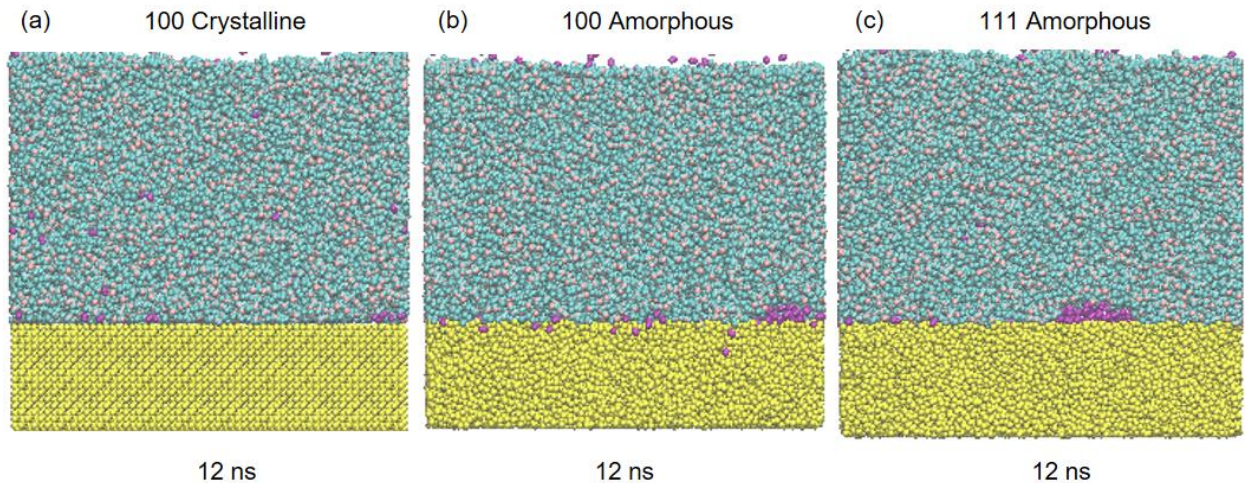

**Figure 7.** MD simulations showing the inability of surface nanobubbles to form on crystalline (a) or amorphous (b, c) Si surfaces even at 1.4% $N_2$ concentration. Simulations were run for 12 ns.

**4. Conclusion**

Experimental and simulation results presented in this work support the concept of top-down positioning of surface nanobubbles *via* FIB patterning. Using a minimal step height (7 nm) to distinguish between nanopatterned regions of the substrate, fluid AFM could identify the presence of alternating lines of 75 nm width displaying the presence and absence of surface nanobubbles. Degassed experiments confirmed the features to be surface nanobubbles. When OTS was applied post-FIB milling, nanobubbles were detected across the whole substrate. These results confirm that nanobubbles were selectively placed on hydrophobic stripes but not on the surrounding hydrophilic trenches due to chemical heterogeneity, and not topography. With pre-designed hydrophobicity, the periodic striped nanostructured surface generated by FIB patterning can effectively regulate the position of surface nanobubbles, with nanoscale control of bubble position.

**References**

(1) Lou, S.-T.; Ouyang, Z.-Q.; Zhang, Y.; Li, X.-J.; Hu, J.; Li, M.-Q.; Yang, F.-J. Nanobubbles on Solid Surface Imaged by Atomic Force Microscopy. *Journal of Vacuum Science & Technology B: Microelectronics and Nanometer Structures Processing, Measurement, and Phenomena* **2000**, *18* (5), 2573–2575. https://doi.org/10.1116/1.1289925.

(2) Ishida, N.; Inoue, T.; Miyahara, M.; Higashitani, K. Nano Bubbles on a Hydrophobic Surface in Water Observed by Tapping-Mode Atomic Force Microscopy. *Langmuir* **2000**, *16* (16), 6377–6380. https://doi.org/10.1021/la000219r.

(3) Zhang, X. H.; Zhang, X. D.; Lou, S. T.; Zhang, Z. X.; Sun, J. L.; Hu, J. Degassing and Temperature Effects on the Formation of Nanobubbles at the Mica/Water Interface. *Langmuir* **2004**, *20* (9), 3813–3815. https://doi.org/10.1021/la0364542.

(4) Ding, S.; Xing, Y.; Zheng, X.; Zhang, Y.; Cao, Y.; Gui, X. New Insights into the Role of Surface Nanobubbles in Bubble-Particle Detachment. *Langmuir* **2020**, *36* (16), 4339–4346. https://doi.org/10.1021/acs.langmuir.0c00359.

(5) Liu, G.; Craig, V. S. J. Improved Cleaning of Hydrophilic Protein-Coated Surfaces Using the Combination of Nanobubbles and SDS. *ACS Appl. Mater. Interfaces* **2009**, *1* (2), 481–487. https://doi.org/10.1021/am800150p.

(6) Wu, Z.; Chen, H.; Dong, Y.; Mao, H.; Sun, J.; Chen, S.; Craig, V. S. J.; Hu, J. Cleaning Using Nanobubbles: Defouling by Electrochemical Generation of Bubbles. *Journal of Colloid and Interface Science* **2008**, *328* (1), 10–14. https://doi.org/10.1016/j.jcis.2008.08.064.

(7) Sobhy, A.; Tao, D. Nanobubble Column Flotation of Fine Coal Particles and Associated Fundamentals. *International Journal of Mineral Processing* **2013**, *124*, 109–116. https://doi.org/10.1016/j.minpro.2013.04.016.

(8) Gao, L.; Ni, G.-X.; Liu, Y.; Liu, B.; Castro Neto, A. H.; Loh, K. P. Face-to-Face Transfer of Wafer-Scale Graphene Films. *Nature* **2014**, *505* (7482), 190–194. https://doi.org/10.1038/nature12763.

(9) Wang, Y.; Bhushan, B. Boundary Slip and Nanobubble Study in Micro/Nanofluidics Using Atomic Force Microscopy. *Soft Matter* **2010**, *6* (1), 29–66. https://doi.org/10.1039/B917017K.

(10) Agrawal, A.; Park, J.; Ryu, D. Y.; Hammond, P. T.; Russell, T. P.; McKinley, G. H. Controlling the Location and Spatial Extent of Nanobubbles Using Hydrophobically Nanopatterned Surfaces. *Nano Lett.* **2005**, *5* (9), 1751–1756. https://doi.org/10.1021/nl051103o.

(11) Nishiyama, T.; Takahashi, K.; Ikuta, T.; Yamada, Y.; Takata, Y. Hydrophilic Domains Enhance Nanobubble Stability. *ChemPhysChem* **2016**, *17* (10), 1500–1504. https://doi.org/10.1002/cphc.201501181.

(12) Belova, V.; Shchukin, D. G.; Gorin, D. A.; Kopyshev, A.; Möhwald, H. A New Approach to Nucleation of Cavitation Bubbles at Chemically Modified Surfaces. *Phys. Chem. Chem. Phys.* **2011**, *13* (17), 8015. https://doi.org/10.1039/c1cp20218a.

(13) Wang, L.; Wang, X.; Wang, L.; Hu, J.; Wang, C. L.; Zhao, B.; Zhang, X.; Tai, R.; He, M.; Chen, L.; Zhang, L. Formation of Surface Nanobubbles on Nanostructured Substrates. *Nanoscale* **2017**, *9* (3), 1078–1086. https://doi.org/10.1039/C6NR06844H.

(14) Wang, Y.; Bhushan, B.; Zhao, X. Nanoindents Produced by Nanobubbles on Ultrathin Polystyrene Films in Water. *Nanotechnology* **2009**, *20* (4), 045301. https://doi.org/10.1088/0957-4484/20/4/045301.

(15) Tarábková, H.; Janda, P. Nanobubble Assisted Nanopatterning Utilized for *Ex Situ* Identification of Surface Nanobubbles. *J. Phys.: Condens. Matter* **2013**, *25* (18), 184001. https://doi.org/10.1088/0953-8984/25/18/184001.

(16) Berkelaar, R. P.; Zandvliet, H. J. W.; Lohse, D. Covering Surface Nanobubbles with a NaCl Nanoblanket. *Langmuir* **2013**, *29* (36), 11337–11343. https://doi.org/10.1021/la402503f.

(17) Darwich, S.; Mougin, K.; Vidal, L.; Gnecco, E.; Haidara, H. Nanobubble and Nanodroplet Template Growth of Particle Nanorings versus Nanoholes in Drying Nanofluids and Polymer Films. *Nanoscale* **2011**, *3* (3), 1211. https://doi.org/10.1039/c0nr00750a.

(18) Li, Y.; Jia, W.-Z.; Song, Y.-Y.; Xia, X.-H. Superhydrophobicity of 3D Porous Copper Films Prepared Using the Hydrogen Bubble Dynamic Template. *Chem. Mater.* **2007**, *19* (23), 5758–5764. https://doi.org/10.1021/cm071738j.

(19) Munir Nayfeh. Manipulation and Patterning of Surfaces (Nanolithography). In *Fundamentals and Applications of Nano Silicon in Plasmonics and Fullerines*; Elsevier, 2018; pp 89–137. https://doi.org/10.1016/B978-0-323-48057-4.00005-0.

(20) Lawson, R. A.; Robinson, A. P. G. Overview of Materials and Processes for Lithography. In *Frontiers of Nanoscience*; Elsevier, 2016; Vol. 11, pp 1–90. https://doi.org/10.1016/B978-0-08-100354-1.00001-6.

(21) Rumyantsev, A. V.; Borgardt, N. I.; Prikhodko, A. S.; Chaplygin, Y. A. Characterizing Interface Structure between Crystalline and Ion Bombarded Silicon by Transmission Electron Microscopy and Molecular Dynamics Simulations. *Applied Surface Science* **2021**, *540*, 148278. https://doi.org/10.1016/j.apsusc.2020.148278.

(22) Satake, S.; Ono, K.; Shibahara, M.; Taniguchi, J. Molecular Dynamics Simulation of Ga+ Ion Collision Process. *Nuclear Instruments and Methods in Physics Research Section B:*

*Beam Interactions with Materials and Atoms* **2013**, *307*, 235–239. https://doi.org/10.1016/j.nimb.2012.12.066.

(23) Ageev, O. A.; Kolomiytsev, A. S.; Konoplev, B. G. Formation of Nanosize Structures on a Silicon Substrate by Method of Focused Ion Beams. *Semiconductors* **2011**, *45* (13), 1709–1712. https://doi.org/10.1134/S1063782611130021.

(24) Ozcelik, H. G.; Ozdemir, A. C.; Kim, B.; Barisik, M. Wetting of Single Crystalline and Amorphous Silicon Surfaces: Effective Range of Intermolecular Forces for Wetting. *Molecular Simulation* **2020**, *46* (3), 224–234. https://doi.org/10.1080/08927022.2019.1690145.

(25) Lehrer, C.; Frey, L.; Petersen, S.; Ryssel, H. Limitations of Focused Ion Beam Nanomachining. *J. Vac. Sci. Technol. B* **2001**, *19* (6), 2533. https://doi.org/10.1116/1.1417553.

(26) Yamada, Y.; Takahashi, K.; Ikuta, T.; Nishiyama, T.; Takata, Y.; Ma, W.; Takahara, A. Tuning Surface Wettability at the Submicron-Scale: Effect of Focused Ion Beam Irradiation on a Self-Assembled Monolayer. *J. Phys. Chem. C* **2016**, *120* (1), 274–280. https://doi.org/10.1021/acs.jpcc.5b09019.

(27) Ada, E. T. Ion Beam Modification and Patterning of Organosilane Self-Assembled Monolayers. *J. Vac. Sci. Technol. B* **1995**, *13* (6), 2189. https://doi.org/10.1116/1.588102.

(28) Rappe, A. K.; Casewit, C. J.; Colwell, K. S.; Goddard, W. A.; Skiff, W. M. UFF, a Full Periodic Table Force Field for Molecular Mechanics and Molecular Dynamics Simulations. *J. Am. Chem. Soc.* **1992**, *114* (25), 10024–10035. https://doi.org/10.1021/ja00051a040.

(29) Hu, K.; Luo, L.; Sun, X.; Li, H. Unraveling the Effects of Gas Species and Surface Wettability on the Morphology of Interfacial Nanobubbles. *Nanoscale Adv.* **2022**, *4* (13), 2893–2901. https://doi.org/10.1039/D2NA00009A.

(30) Zou, Z.-L.; Quan, N.-N.; Wang, X.-Y.; Wang, S.; Zhou, L.-M.; Hu, J.; Zhang, L.-J.; Dong, Y.-M. The Properties of Surface Nanobubbles Formed on Different Substrates. *Chinese Phys. B* **2018**, *27* (8), 086803. https://doi.org/10.1088/1674-1056/27/8/086803.

(31) Simonsen, A. C.; Hansen, P. L.; Klösgen, B. Nanobubbles Give Evidence of Incomplete Wetting at a Hydrophobic Interface. *Journal of Colloid and Interface Science* **2004**, *273* (1), 291–299. https://doi.org/10.1016/j.jcis.2003.12.035.

(32) Zhang, X. H.; Maeda, N.; Craig, V. S. J. Physical Properties of Nanobubbles on Hydrophobic Surfaces in Water and Aqueous Solutions. *Langmuir* **2006**, *22* (11), 5025–5035. https://doi.org/10.1021/la0601814.

(33) Lohse, D.; Zhang, X. Surface Nanobubbles and Nanodroplets. *Rev. Mod. Phys.* **2015**, *87* (3), 981–1035. https://doi.org/10.1103/RevModPhys.87.981.

(34) Lohse, D.; Zhang, X. Pinning and Gas Oversaturation Imply Stable Single Surface Nanobubbles. *Phys. Rev. E* **2015**, *91* (3), 031003. https://doi.org/10.1103/PhysRevE.91.031003.

(35) An, H.; Tan, B. H.; Zeng, Q.; Ohl, C.-D. Stability of Nanobubbles Formed at the Interface between Cold Water and Hot Highly Oriented Pyrolytic Graphite. *Langmuir* **2016**, *32* (43), 11212–11220. https://doi.org/10.1021/acs.langmuir.6b01531.